\documentclass[10pt,twocolumn]{article}
\usepackage[columnsep=0.4in]{geometry}
\usepackage{lineno}

\usepackage{url}
\usepackage{amsmath}
\usepackage{amssymb}
\usepackage[ruled]{algorithm2e}
\usepackage{graphicx} 

\newcommand{\norm}[1]{\|#1\|}

\newcommand{\T}{^T\!}

\newcommand{\pmat}[1]{\begin{pmatrix}#1\end{pmatrix}} 

\newcommand{\B}[2]{{\footnotesize\bf#1#2}} 
\newcommand{\Bzz}{{\footnotesize\bf--}} 
\newcommand{\E}[1]{\hbox{e\raisebox{1pt}{\tiny$#1$}}}    
\newcommand{\Meszaros}{M\'esz\'aros}
\newcommand{\mathone}{\mathbf{1}}

\begin{document}


\title{\textsf{Reliable and efficient solution of genome-scale models of}
    \\ \textsf{Metabolism and macromolecular Expression}}

\author{\sf
        Ding Ma%
        \footnote{\textsf{Dept of Management Science and Engineering,
                        Stanford University, Stanford, CA, USA.
                  Correspondence should be addressed to M.~S.\ 
                  (saunders@stanford.edu).}}\;,
        Laurence Yang%
        \footnote{\textsf{Dept of Bioengineering, University of California,
                        San Diego, CA, USA.}}\;,    
        Ronan M. T. Fleming%
        \footnote{\textsf{Luxembourg Centre for Systems Biomedicine,
                        University of Luxembourg, Luxembourg.}}\;,
        Ines Thiele$^{\ddagger}$,
     \\ \sf
        Bernhard O. Palsson$^{\dagger}$%
        \footnote{\textsf{Novo Nordisk Foundation Center for Biosustainability,
                          The Technical University of Denmark,
                          2970 H{\o}rsholm, Denmark.}}
        \ \&\ 
        Michael A. Saunders$^*$%
       }
\date{\textsf{\today}}

\twocolumn

\maketitle

\vspace*{-22pt}

\noindent
\begin{flushleft}%
\textsf{\textbf{Constraint-Based Reconstruction and Analysis (COBRA) is currently the
  only methodology that permits integrated modeling of Metabolism and
  macromolecular Expression (ME) at genome-scale.  Linear optimization
  computes steady-state flux solutions to ME models, but flux values
  are spread over many orders of magnitude.  Standard double-precision
  solvers may return inaccurate solutions or report that no solution
  exists.  Exact simplex solvers are extremely slow and hence not practical
  for ME models 
  that currently have 70,000 constraints and variables and
  will grow larger.  We
  have developed a quadruple-precision version of our linear and
  nonlinear optimizer MINOS, and a solution procedure (DQQ) involving
  Double and Quad MINOS that achieves efficiency and reliability for
  ME models.
  DQQ enables extensive use of large, multiscale, linear and nonlinear
  models in systems biology and many other applications.
}}

\end{flushleft}


\medskip

\noindent
Constraint-Based Reconstruction and Analysis (COBRA)
\cite{Palsson2006} has been applied successfully to predict phenotypes
for a range of genome-scale biochemical processes. The popularity of
COBRA is partly due to the efficiency of the underlying optimization
algorithms, permitting genome-scale modeling at a particular timescale
using readily available open source software
\cite{schellenberger2011quantitative,COBRApy} and industrial quality
optimization algorithms \cite{Gurobi,CPLEX,MOSEK}.  A widespread
application of COBRA is the modeling of steady states in genome-scale
Metabolic models (M models).  COBRA has also been used to model steady
states in macromolecular Expression networks (E models), which
stoichiometrically represent the transcription, translation,
post-translational modification and formation of all protein complexes
required for macromolecular biosynthesis and metabolic reaction
catalysis \cite{thiele2009gcr, Thiele:2009d}. COBRA of metabolic
networks or expression networks depends on 
numerical optimization algorithms to compute solutions to certain
model equations, or to determine that no solution exists.  Our purpose
is to discuss available options and to demonstrate an approach that is
reliable and efficient for ever larger networks.

Metabolism and macromolecular Expression (ME) models
 have opened a whole new vista for predictive mechanistic
modeling of cellular processes, but their size and multiscale nature
pose a challenge to standard linear optimization (LO) solvers based on 16-digit double-precision
floating-point arithmetic. Standard LO solvers usually apply scaling techniques
\cite{Fourer82a,tomlin1975scaling}
to problems that are not already well scaled.
The scaled problem typically solves more efficiently and accurately,
but the solver must then unscale the solution, and this may generate
significant primal or dual infeasibilities in the original problem
(the constraints or optimality conditions may not be accurately
satisfied).  

A \emph{lifting} approach \cite{SunFTS2013} has been implemented
to alleviate this difficulty with multiscale problems.
Lifting reduces the largest matrix entries by introducing auxiliary
constraints and variables.  This approach has permitted standard (double-precision)
LO solvers to find more accurate solutions, even though the final objective value is still not satisfactory
(Table~\ref{tab:MEobj}).
Another approach to increasing the precision is to use an exact
solver.  An exact simplex solver QSopt\_ex
\cite{Applegate2007,QSopt_ex} has been used for a ME model of
\emph{Thermotoga maritima} \cite{Josh2012} (model TMA\_ME)
representing a network with about 18,000 metabolites and reactions.
The solution time 
was about two
weeks, compared to a few minutes (Table~\ref{tab:MEtime}) for a standard double-precision
solver, but the latter's final objective value had only
one correct digit (Table~\ref{tab:MEobj}).
QSopt\_ex has since been applied to a collection of 98 metabolic
models by Chindelvitch et al.\ \cite{Mongoose2014}
via their MONGOOSE toolbox. 
Most of the
98 models have less than 1000 metabolites and reactions.  QSopt\_ex
required about a day to solve all models \cite{Mongoose2014},
compared to a few seconds in total for a standard solver. 

To advance COBRA for increasingly large
biochemical networks, solvers that perform more efficiently than exact solvers 
and also perform more reliably than standard LO solvers are definitely needed. 
%
Gleixner et al.\ \cite{IterativeRefinement2015,gleixner2015thesis}
have addressed this need, and Chapter 4 of \cite{gleixner2015thesis}
is devoted to multiscale metabolic networks, showing significant
improvement relative to CPLEX \cite{CPLEX}.  Our work is
complementary and confirms the value of enhancing the simplex
solver in \cite{IterativeRefinement2015,gleixner2015thesis}
to employ quadruple-precision computation, as we have done here.

Let Single/Double/Quad denote the main floating-point options,
with about 7, 16, and 34 digits of precision respectively.  For many
years, scientific computation has advanced in two complementary ways:
improved \emph{algorithms} and improved \emph{hardware}. 
Compilers have typically evaluated expressions using the same arithmetic as the variables'
data type. Most scientific codes apply Double
variables and Double arithmetic throughout (16 significant digits
stored in 64-bit words). The floating-point hardware often has
\emph{slightly extended precision} (80-bit registers). 
Kahan \cite{Kahan2011}
notes that early C compilers generated Double instructions
for all floating-point computation \emph{even for program variables
  stored in single precision}. 
Thus for a brief period, C programs were
serendipitously more reliable than typical Fortran programs of the
time. 
(For Single variables $a$ and
$b$, Fortran compilers would use Single
arithmetic to evaluate the basic expressions $a\pm b$, $a{*}b$, $a/b$,
whereas C compilers would transfer $a$ and $b$ to longer
registers and operate on them using Double arithmetic.) Most
often, the C compiler's extra precision was not needed, but
occasionally it did make a critical difference. 
Kahan calls this the
\emph{humane} approach to debugging complex numerical software.
Unfortunately, Quad hardware remains very rare
and for the foreseeable future will be simulated on most machines
by much slower software.
Nevertheless, 
we believe the time has come to produce Quad versions of key
sparse-matrix packages and large-scale optimization solvers 
for multiscale problems. 

Here, we report the development and biological application of Quad
MINOS, a quadruple-precision version of our general-purpose,
industrial-strength linear and nonlinear optimization solver MINOS
\cite{MurS78,MurS82}.  We also developed a Double-Quad-Quad MINOS
procedure (DQQ) that combines the use of Double and Quad 
solvers in order to achieve a balance between efficiency in
computation and accuracy of the solution. We extensively tested this
DQQ procedure on 83 genome-scale metabolic network
models (M models) obtained from the UCSD Systems Biology repository
\cite{UCSD,MultiscaleCollaboration}
and 78 from the BiGG database \cite{KingLu2015}. 
We also applied DQQ to ME models of
\emph{Thermotoga maritima} \cite{Josh2012} (about 18,000 metabolites
and reactions) and \emph{E. coli} K12 MG1655 \cite{Ines2012ME} (about
70,000 metabolites and reactions).
For M models, we find that
Double MINOS alone is sufficient to obtain non-zero
steady-state solutions that satisfy feasiblility and optimality
conditions with a tolerance of $10^{-7}$. For ME models, application
of our DQQ procedure resulted in non-zero steady-state solutions that
satisfy feasibility and optimality conditions with a tolerance of
$10^{-20}$.  The largest model, a lifted version of the \emph{E. coli}
ME model, required 4.5 hours, 
while an exact solver would take months. 



Thus, we expect our DQQ procedure to be a robust and efficient tool for
the increasingly detailed study of biological processes, such as
metabolism and macromolecular synthesis, and many other scientific
fields.

\subsection*{\textsf{\textbf{Results}}}
\vspace{-6pt}
\noindent
\textbf{Efficient combination of Double and Quad.}
To achieve reliability and efficiency on multiscale problems, in general form as \emph{linear
 optimization} (LO):
\begin{eqnarray} \label{eq:fba}
  \min\ c\T v \text{\ \ s.t.\ \ } Sv=0, \ \ \ell \le v \le u,
\end{eqnarray}
we developed the following 3-step procedure. 



\smallskip
\noindent
\textbf{DQQ procedure.}  

\textit{Step D} (Cold start in Double with scaling):
  Apply Double MINOS with moderately strict options.
  Save a final basis file.
  
\textit{Step Q1} (Warm start in Quad with scaling):
  Start Quad MINOS from the saved file with stricter
  options.  Save a final basis file.
  
\textit{Step Q2} (Warm start in Quad without scaling):
  Start Quad MINOS from the second saved file with no scaling but
  stricter LU options.

\smallskip

 DQQ is described further in Algorithm \ref{alg:DQQ}, where 
 loop \textbf{\footnotesize 1} is the primal simplex method,
 $P$ is a permutation matrix, and $\delta_1$, $\delta_2$ are
 Feasibility and Optimality tolerances.  MINOS
 terminates loop \textbf{\footnotesize 1} when the (possibly scaled)
 bounds on $v$ are satisfied to within $\delta_1$ and $z_j/(1 +
 \norm{y}_\infty)$ has the correct sign to within $\delta_2$.
 Table~\ref{tab:options} shows the default runtime options for Double
 MINOS and the options chosen for each step of DQQ.
Scale specifies whether the problem data should be scaled before the
problem is solved (and unscaled after).  Tolerances $\delta_1$,
$\delta_2$ specify how well the primal and dual constraints of the
(possibly scaled) problem should be satisfied.  Expand frequency
controls the MINOS anti-degeneracy procedure \cite{GilMSW89c}.  The LU
tolerances balance stability and sparsity when LU factors of $B$
are computed or updated.

 \begin{algorithm}[!t]
 	\KwData{Linear program \eqref{eq:fba}} 
 	\KwResult{Flux vector $v^*$, basis partition $SP\!=\!(B\ N)$,
          one of three states: optimal, infeasible, or unbounded
          (possible if infinite $\ell_j$, $u_j$ exist)}

        \smallskip

 	\textbf{\textit{Step D}}: use Double MINOS with scaling\;
    
     \nl\Repeat{$\forall j \in N$, $z_j \leq \tau$ if $v_j = \ell_j$, and
                $z_j \geq - \tau$ if $v_j = u_j$ (optimal);
                or fail to find $\ell - \delta_1 \le v \le u + \delta_1$ (infeasible);
                or fail to find a new $B$ (unbounded)}
       {Find a nonsingular basis matrix $B$ from the columns of $S$
       so that $SP = (B\ N)$\;
       Find $v = P(v_B, v_N)$ with $Sv \equiv Bv_B + Nv_N = 0$\;
 	  Partition $c$ accordingly as $c = P(c_B, c_N)$\;
       Solve $B\T y = c_B$\;
       Set $z_N  \leftarrow c_N - N\T y$; \ 
       $\tau \leftarrow (1+\norm{y}_\infty) \delta_2$\;
     }
    

     \smallskip

     \textbf{\textit{Step Q1}}: use Quad MINOS with scaling\; Start with the saved $B$ from
     \textit{Step D} to run loop \textbf{\footnotesize 1} 
     to find a new $B$\;

     \smallskip

     \textbf{\textit{Step Q2}}: use Quad MINOS without scaling\; Start with the saved $B$ from
     \textit{Step Q1} to run loop \textbf{\footnotesize 1} 
     to reach a final $B$\;
     \caption{\small DQQ procedure \label{alg:DQQ}}
 \end{algorithm}


\begin{table}[!t] 
  \vspace*{-4pt}
  \caption{\small MINOS runtime options (defaults and those selected for each
    step of the DQQ procedure).}
\label{tab:options}
\vspace{-4pt}

\begin{center}
\begin{footnotesize}
\begin{tabular}{@{\,}l@{\quad}|r@{\quad}r@{\quad}r@{\quad}r@{\,}}
\hline                        & Default  &  Step D  & Step Q1 & Step Q2
\\ \hline
  Precision                  & Double   &  Double  &   Quad  &   Quad 
\\
   Scale                      &     Yes  &     Yes  &    Yes  &     No 
\\ Feasibility tol $\delta_1$ &    1e-6  &    1e-7  &  1e-15  &  1e-15 
\\ Optimality  tol $\delta_2$ &    1e-6  &    1e-7  &  1e-15  &  1e-15 
\\ Expand frequency           &   10000  &  100000  & 100000  & 100000
\\ LU Factor tol              &   100.0  &     1.9  &   10.0  &    5.0 
\\ LU Update tol              &    10.0  &     1.9  &   10.0  &    5.0 
\\ \hline
\end{tabular}
\end{footnotesize}
\end{center}
\end{table}

Steps D and Q1 are usually sufficient, but if Q1 is interrupted, Q2
provides some insurance and ensures that
the 
tolerances $\delta_1$ and $\delta_2$ are imposed upon the original
problem (not the scaled problem).  For conventional Double solvers, it
is reasonable to set tolerances in the range $10^{-6}$ to $10^{-8}$.  For
Quad MINOS, we set $\delta_1 = \delta_2 = 10^{-15}$ to be sure of
capturing 
reaction fluxes $v_j$ as small as $O(10^{-10})$.


\textbf{Small M models.}  Of the
98 metabolic network models in the
UCSD Systems Biology repository \cite{UCSD}, A. Ebrahim
was able to parse 83 models \cite{FBA83generate} and compute solutions
with a range of solvers \cite{FBA83solve}.  We constructed MPS files
for the 83 models \cite{MultiscaleCollaboration} and solved them via
DQQ.  Most models have less than 1000 metabolites and reactions.
Almost all models solved in less than 0.08 seconds, and many in less
than 0.01 seconds.  The total time was less than 3 seconds.
In contrast, the exact arithmetic solver needs a day \cite{Mongoose2014}.


\textbf{Large ME models.} COBRA can be used to
stoichiometrically couple metabolic and macromolecular expression
networks with single nucleotide resolution at genome-scale
\cite{Ines2012ME, Josh2012}. The corresponding Metabolic and
macromolecular Expression models (ME models) explicitly represent
catalysis by macromolecules, and in turn, metabolites are substrates
in macromolecular synthesis reactions. 
These reconstructions lead to the first
multi-timescale and genome-scale stoichiometric models, as they
account for multiple cellular functions operating on widely different
timescales and typically account for about 40 percent of a
prokaryote's open reading frames. A typical M model might be
represented by 1000 reactions generated by hand
\cite{feist2007gsm}. In contrast, ME models can have more than 50,000
reactions, most of which have been generated algorithmically from
template reactions (defined in the literature) and omics data
\cite{Ines2012ME, Josh2012}.
Typical net metabolic reaction rates are 6 orders of magnitude faster
than macromolecular synthesis reaction rates (millimole/gDW vs
nanomole/gDW, gDW = gram dry weight), and the number of metabolic
moieties in a macromolecule can be many orders of magnitude larger
than in a typical metabolite. The combined effect is that the
corresponding ME models have biochemically significant digits over
many orders of magnitude.  
When Flux Balance Analysis (FBA) is augmented with coupling constraints
\cite{Ines2010AOS} that constrain the ratio between catalytic usage of
a molecule and synthesis of the same molecule, the corresponding linear
optimization
problem is multiscale in the sense that both data values and solution
values have greatly varying magnitudes.  For a typical ME model, input
data values (objective, stoichiometric or coupling coefficients, or
bounds) differ by 
6 orders of magnitude, and biochemically meaningful solution values
can be as large as $10^8$ or as small as $10^{-10}$.
%


The results of DQQ on three large ME models
TMA\_ME, GlcAerWT, and GlcAlift are shown in
Tables~\ref{tab:MEtime}--\ref{tab:MEobj}, including problem dimensions
($m, n$), number of nonzero entries (nnz$(S)$), norms of the optimal
primal and dual solution vectors $(v^*\!$, $y^*)$, number of
iterations, runtime, objective value, primal and dual infeasibility
after each step (Pinf and Dinf), and total solve time for each model.
(The constraints in \eqref{eq:fba}
are satisfied to
within Pinf, and $z_j/(1 + \norm{y^*}_\infty)$ has the correct sign to
within Dinf, where $B^T y = c_B$ for the optimal basis $B$, and $z = c
- S^T y$.)

\textit{\textbf{TMA\_ME}} developed by Lerman et al.\ \cite{Josh2012}
has some large matrix entries $|S_{ij}|$ and many small solution
values $v_j$ that are meaningful to systems
biologists.  
%
For example, transcription and translation rates can have values
$O(10^{-10})$ or less, which is much smaller than metabolic reactions.
These small values are linked to large matrix entries arising from
building large macromolecules from smaller constituents
\cite{Ines2012ME}.
The ME part of the model also contains small matrix entries.
For instance, enzyme levels are estimated in ME models by dividing
certain metabolic fluxes by ``effective rate constants.'' Because
these constants are typically large (e.g., 234,000 h$^{-1}$), the
matrix entries (the inverse of the rate constants) become small.
In step D, almost all iterations went on finding a feasible solution,
and the objective then had the correct order of magnitude
(but only one correct digit).
Step Q1 improved the accuracy, and Step Q2 provided confirmation. 
Note that the efficiency advantage of our approach is also evident: 385
seconds solve time for DQQ (Total time in Table~\ref{tab:MEtime})
compared to 2 weeks using exact arithmetic \cite{Josh2012}.

\begin{table}[!t] 
  \caption{\small Three large ME biochemical network models
    TMA\_ME, GlcAerWT, GlcAlift 
    \cite{Josh2012, Ines2012ME, SunFTS2013}.
    Dimensions of $m \times n$ constraint matrices $S$, 
    size of the largest optimal primal and dual variables
    $v^*\!$, $y^*$, number of iterations and runtimes in seconds for
    each step, and the total runtime of each model.}
\label{tab:MEtime}
\medskip
\centering
\begin{footnotesize}
\begin{tabular}{|l|r|r|r|}
\hline
ME model & TMA\_ME & GlcAerWT & GlcAlift
\\ \hline
 $m$ & 18210 & 68300 & 69529 \\
 $n$ & 17535 & 76664 & 77893 \\
  nnz($S$) &  336302 &   926357 &   928815 \\
     $\max |S_{ij}|$ & 2.1\E+04 & 8.0\E+05 & 2.6\E+05 \\
 $\norm{v^*}_\infty$ & 5.9\E+00 & 6.3\E+07 & 6.3\E+07 \\
 $\norm{y^*}_\infty$ & 1.1\E+00 & 2.4\E+07 & 2.4\E+07 \\
  D itns   &   21026 &    47718 &    93857 \\
  D time   &   350.9 &  10567.8 &  15913.7 \\
  Q1 itns  &     597 &     4287 &     1631 \\
  Q1 time  &    29.0 &   1958.9 &    277.3 \\
  Q2 itns  &       0 &        4 &        1 \\
  Q2 time  &     5.4 &     72.1 &     44.0 \\
  Total time &   385 &    12599 &    16235 \\         
\hline
\end{tabular}
\end{footnotesize}
\vspace*{-8pt}
\end{table}

\begin{table} 
  \caption{\small Three large ME biochemical network models
    TMA\_ME,  GlcAerWT, GlcAlift 
    \cite{Josh2012, Ines2012ME, SunFTS2013}.
           Optimal objective value of each step, 
           Pinf and Dinf = final maximum primal and dual infeasibilities
           ($\log_{10}$ values tabulated, except \Bzz{} means 0).
           Bold figures show the final \textit{(step Q2)} Pinf and Dinf.}
\label{tab:MEobj}
\medskip
\centering
\begin{footnotesize}
\hspace*{0.6pt}%
\begin{tabular}{|l|l|r|r|r|}
   \hline
   ME model &Step& Objective\qquad\qquad & Pinf & Dinf
\\ \hline
   TMA\_ME  & D  &   8.3789966820\E-07 & $-$06   & $-$05
\\          & Q1 &   8.7036315385\E-07 & $-$25   & $-$32
\\          & Q2 &   8.7036315385\E-07 & \Bzz    & $-$\B32
\\ GlcAerWT & D  &$-$6.7687059922\E+05 & $-$04   & $+$00
\\          & Q1 &$-$7.0382449681\E+05 & $-$07   & $-$26
\\          & Q2 &$-$7.0382449681\E+05 & $-$\B21 & $-$\B22
\\ GlcAlift & D  &$-$5.3319574961\E+05 & $-$03   & $-$01
\\          & Q1 &$-$7.0434008750\E+05 & $-$08   & $-$22
\\          & Q2 &$-$7.0434008750\E+05 & $-$\B18 & $-$\B23
\\ \hline
\end{tabular}
\end{footnotesize}
\vspace*{-8pt}
\end{table}

Two slightly different versions of this model provided welcome
empirical evidence that the optimal objective and solution values
do not change significantly when the problem data are perturbed
by $O(10^{-6})$ (see Supplementary Information).

\textit{\textbf{GlcAerWT}} is a ME model from the detailed study by
Thiele et al.\
\cite{Ines2012ME}.  
After 33,000 iterations, Double MINOS began to report singularities
following updates to the basis LU factors (71 times during the next
15,000 iterations).  After 47,718 iterations (D itns in
Table~\ref{tab:MEtime}), step~D terminated with maximum
primal and dual infeasibilities $O(10^{-4})$ and $O(1)$
(Pinf and Dinf in Table~\ref{tab:MEobj}).  These were small enough
to be classified ``Optimal'', but we see that the final objective
value $-6.7687\E+05$ had no correct digits compared to
$-7.0382\E+05$ in steps Q1 and Q2.
For large models, step Q1 is 
important.  It required significant work:
4,287 iterations costing 1958.9 seconds (Q1 itns and time in
Table~\ref{tab:MEtime}).
%
Step Q2 quickly confirmed the final objective value with high accuracy.
This, the largest ME model so far,
solved in 12,599 seconds (3.5 hours) compared to an expected
time of months for an exact solver. 


\textit{\textbf{GlcAlift}} is motivated by the difficulties with
solving TMA\_ME and GlcAerWT in Double arithmetic.  The lifting
technique of \cite{SunFTS2013} 
was applied to GlcAerWT to reduce some of the large matrix values.
The aim of lifting is to remove the need for scaling (and hence
the difficulties with unscaling), but with DQQ we do scale
in step D because steps Q1 and Q2 follow.  Our experience is that
lifting improves accuracy for Double solvers but substantially
increases the simplex iterations.  On GlcAlift, Double MINOS again
reported frequent singularities following basis updates (235 times
starting near iteration 40,000).  It took 93,857 iterations (D itns in
Table~\ref{tab:MEtime}), twice as many as GlcAerWT, with only a slight
improvement in $\max \{\mathrm{Pinf},\mathrm{Dinf}\}$
(Table~\ref{tab:MEobj}).  Double MINOS with scaling on the lifted model 
couldn't reach agreement with the final
objective $-$7.0434008750\E+05 in steps Q1 and Q2, and the total solve
time increased (4.5 hours), mostly in step~D. 
The objective function for both GlcA models is to maximize variable
$v_{60069}$.  
%
%
The fact that the step~D objective values have no
correct digits 
illustrates the challenge these models present.
Starting from
the basis that the Double solver reaches, steps Q1 and Q2 are accurate
and efficient.  Theoretically, the Q2 objectives for GlcAerWt and
GlcAlift should agree, but limited precision in the data files
could explain why there is just 3-digit agreement. 
 
The Tomlab interface \cite{TOMLAB} and CPLEX were 
used by Thiele et al.\ \cite{Ines2012ME} to improve the results for 
standard Double solvers. 
On the NEOS server \cite{NEOS}, Gurobi was unable to solve
GlcAerWT with default parameters (numeric error after nearly 600,000
iterations). 
It performed considerably better on GlcAlift (about 46,000 iterations)
but terminated with a warning of unscaled primal/dual residuals
1.07 and 1.22\E-06.
As shown above, our DQQ procedure 
saves researchers' effort on lifting the model, and is able to 
solve the original model faster (3.5 hours vs 4.5 hours).

Further tests of the DQQ procedure on challenging LO problems are
reported in \textsf{\textbf{Methods}}.  As for the ME models, the simplex method in Double
MINOS usually gives a good starting point for the same simplex method
in Quad MINOS.  Hence, much of the work can be performed efficiently
with conventional 16-digit floating-point hardware to obtain
near-optimal solutions.  For Quad MINOS, 34-digit floating-point
operations are implemented in the compiler's Quad math library via
software (on today's machines).  Each simplex iteration is therefore
considerably slower than by hardware, but the reward is extremely high accuracy.  Of
significant interest is that Quad MINOS almost invariably achieves
\emph{far more accurate solutions than requested} (see bold figures in
Tables~\ref{tab:MEobj} and~\ref{tab:results}).
This is 
a favorable and promising empirical finding.

\subsection*{\textsf{\textbf{Discussion}}}
\vspace{-6pt}
%
Exact solvers compute exact solutions to LO problems
involving rational data.
Although stoi\-chiometric coefficients for chemical reactions
are in principle integers, most genome-scale
metabolic models have non-integer coefficients where
the stoichi\-ometry is known to only a few digits,
e.g., a coefficient in a biomass reaction. Such a stoichiometric
coefficient should not be considered exact data (to be converted into
a rational number for use with an exact solver).
This casts doubt on any effort to
compute an exact solution for a particular FBA problem.


Exact solvers are based on rational arithmetic.  There has been
considerable work on their 
application to important
problems \cite{Koch2003final, Applegate2007, QSopt_ex, Josh2012}.
The use of quadruple-precision and variable-precision floating-point
has also been mentioned \cite{Koch2003final, Applegate2007}.  Here, we
exploit Quad precision more fully on a range of larger
problems, knowing that
current genome-scale models will continue to grow even larger.

While today's advanced LO solvers, such as CPLEX, Gurobi, Mosek, and
Xpress \cite{CPLEX,MOSEK,Gurobi,Xpress}, are effective on a wide range
of large and challenging linear (and mixed integer) optimization
models, the study by Thiele et al.\ \cite{Ines2012ME} emphasizes the
need for improved reliability in solving FBA and ME models in systems
biology.
%
%
Our DQQ procedure has demonstrated that warm starts with Quad solvers
are efficient, and that the accuracy achieved
exceeds requirements by a very safe margin.  Kahan \cite{Kahan2011}
notes that ``\emph{carrying somewhat more precision in the arithmetic
  than twice the precision carried in the data and available for the
  result will vastly reduce embarrassment due to roundoff-induced
  anomalies}'' and that ``\emph{default evaluation in Quad is the
  humane option}.''  The ``humane'' approach---use of Quad
solvers---is certainly more efficient than applying exact solvers.

An intriguing question remains concerning the bold figures in Tables
\ref{tab:MEobj} and \ref{tab:results}.  The primal and dual solutions
obtained with Quad precision are {substantially more accurate
  than the $10^{-15}$ requested}.  The same has been true for all of
the classic set of Netlib problems \cite{lpdata} that we have run.
Kahan \cite{Kahan2011} explains that perturbations get amplified by
singularities near the data.  He describes a \emph{pejorative surface}
of data points where singularity exists, and expects loss of accuracy
as data approaches the surface.  The volume surrounding the pejorative
surface is the danger zone, but: ``\emph{Arithmetic precision is
  usually extravagant enough if it is somewhat more than twice as
  [great] as the data's and the desired result's. Often that shrunken
  volume contains no data}.''  We surmise that Kahan has anticipated
our observed situation, wherein LO problems defined with
double-precision data appear unlikely to be too ill-conditioned for a
Quad solver.

It should be said that exact simplex solvers can also be warm-started,
as noted by Gleixner et al.\ \cite{IterativeRefinement2015,gleixner2015thesis}.
For many models, most of the work could be done by a conventional
Double solver as in our DQQ procedure, and Step Q1 could be replaced
by a call to an exact solver.  
However,
for the GlcA problems, we see in Table~\ref{tab:MEtime} that step Q1
performs a significant number of iterations.  
Thus, warm-starting
an exact solver on large models could remain too expensive to be practical.

Looking ahead, we note that metabolic reconstructions of the form
\eqref{eq:fba} may need to be processed before they can be treated
as stoichiometrically consistent models.  As discussed in
\cite{FlemingVTS2015}, certain rows of $S$ may need to be deleted
according to the solution $\ell$ of the problem
    $\max \norm{\ell}_0$ s.t.\ $S^T\! \ell = 0$, $\ell \ge 0$.
This problem can be approximated by the linear problem
\vspace{-5pt}
\begin{eqnarray}
  \underset{z,\,\ell}{\max} &\!\!\!\!& \mathone^T\! z \nonumber
\\[-5pt] \text{s.t.}        &\!\!\!\!& S^T\! \ell=0,  \hspace*{25pt}  z \leq \ell,
                                                      \label{eq:maxCardAApprox}
\\[-3pt]                    &\!\!\!\!& 0 \le z \le \mathone\alpha,
                             \quad  0 \le \ell \le \mathone\beta, \nonumber
\end{eqnarray}
where scalars $\alpha, \beta$ are proportional to
the smallest molecular mass considered non-zero and
the largest molecular mass allowed
(e.g., $\alpha = 10^{-4}$, $\beta = 10^{4}$).
Note that problem \eqref{eq:maxCardAApprox} involves $S^T$
and is larger than the FBA problem \eqref{eq:fba} itself.
We could not design consistent FBA models in this way
unless we were sure of being able to solve \eqref{eq:maxCardAApprox}
effectively.  Our work here offers assurance of such capability.

We believe that quadruple-precision solutions are now practical for
multiscale applications such as FBA and flux variability analysis
(FVA) computations for ME models in systems biology
\cite{Palsson2006,orthTP2010flux, Ines2012ME, Ines2010fva,
  Ines2010AOS}, and that our DQQ procedure justifies increased
confidence as systems
biologists build ever-larger models to explore new hypotheses about
metabolism and macromolecular synthesis.  Our 
combined use of Double and Quad solvers will lead to solutions
of exceptional accuracy in other areas of computational science
involving multiscale optimization problems. For example, Dattorro
\cite{Dattorro-2015} has derived an approach to analog filter design
that requires a Quad linear or nonlinear solver to deal with a wide
range of frequencies (which must be raised to high powers).  This
application, like ME models with nonlinear constraints
\eqref{eq:nonlinear2}, can be treated with Quad precision and binary
search on a sequence of problems.  We have also treated the nonlinear
constraints directly with the nonlinear algorithms in Quad MINOS
\cite{MurS82,15Yang}.


\subsection*{\textsf{\textbf{Methods}}}
\vspace{-6pt}
\textbf{Multiscale constraint-based modeling.} Consider a network of
biochemical reactions, represented by a stoichiometric matrix $S \in
\mathbb{R}^{m \times n}$ with each row and column corresponding to a
molecular species and biochemical reaction, respectively. $S_{ij}$
respresents the \emph{stoichiometry} of molecular species $i$
participating as a substrate (negative) or product (positive) in
reaction $j$. The evolution of molecular species concentrations with
respect to time ($t$) is given by the ordinary differential
equation
\begin{eqnarray}
   \frac{dx(t)}{dt} = Sv(x(t)),
   \label{eq:governing-eq}
\end{eqnarray}
where $x(t) \in \mathbb{R}^m_{\ge0}$ is a vector of time-dependent
concentrations and $v(x(t)): \mathbb{R}^m_{\ge0} \rightarrow
\mathbb{R}^n$ is a nonlinear function of concentrations, with a form
that depends on the kinetic mechanism of each reaction.

If one assumes that species concentrations are time-invariant, then
the set of all steady-state reaction rates, satisfying $Sv(x)=0$,
may be approximated by the linear \emph{steady-state constraint} $Sv =
0$, where $v \in \mathbb{R}^n$ is a vector of reaction
fluxes. Thermodynamic principles and experimental data can also be
used to specify lower and upper \emph{bound constraints} on reaction
fluxes $\ell \le v \le u$. Biochemical relationships between the rates
of macromolecular synthesis and utilization can be approximated by
coupling of the corresponding reaction fluxes \cite{Ines2010AOS},
e.g., pyruvate kinase reaction flux and the synthesis flux of pyruvate
kinase in a ME model \cite{Ines2012ME}. Flux coupling can be
represented by bounding the ratio between two reaction fluxes with two
coupling coefficients:
\begin{eqnarray}
   \sigma_{\min} \le \frac{v_i}{v_j} \le \sigma_{\max}, \label{eq:couple1}
\end{eqnarray}
where $v_i$ and $v_j$ are a pair of non-negative fluxes. This
nonlinear constraint can be reformulated into a pair of linear
\emph{coupling constraints}
\begin{eqnarray}
   \sigma_{\min} v_j \le v_i, \quad v_i \le \sigma_{\max} v_j,
\label{eq:couple2}
\end{eqnarray}
or more generally a set of linear inequalities $Cv\le d$. In addition
to the aforementioned physicochemical and biochemical contraints, one
may hypothesize a biologically motivated objective.  For example, in
modeling a growing cell, one may hypothesize that the objective is to
maximize the rate of a biomass synthesis reaction.  Typically, a
biomass synthesis reaction is created with experimentally determined
stoichiometric coefficients, each of which represents the relative
composition of a cellular biomass constituent.  
Optimization of a
linear combination of reaction fluxes $c\T v$
leads to linear optimization problems: \eqref{eq:fba}.
Flux balance analysis of a ME model with coupling constraints results
in an ill-scaled instance of this 
problem because the stoichiometric coefficients and coupling coefficients vary
over many orders of magnitude.



%
%
%
%
%

\medskip
\noindent
\textbf{MINOS implementation.}  
MINOS \cite{MurS78, MurS82} is a linear and nonlinear optimization
solver implemented in Fortran 77 to solve problems of the form
\begin{eqnarray}
   \min_v\ c\T v + \varphi(v)
   \text{\ \ s.t.\ \ } \ell \le \pmat{v \\ Sv \\ f(v)} \le u,
   \label{eq:NLP}
\end{eqnarray} 
%
where $\varphi(v)$ is a smooth nonlinear function
and $f(v)$ is a vector of smooth nonlinear functions
(see Supplementary Information).

\medskip
\noindent
\textbf{Further tests of DQQ.}
We report results from
the primal simplex solver in Double MINOS and Quad MINOS
on two sets of challenging LO problems
shown in Table~\ref{tab:dims}.  As with the M and ME models, all
runs were on a 2.93 GHz Apple iMac with quad-core Intel i7, using the
gfortran compiler with -O flag (GNU Fortran 5.2.0).
The problems were input from files in the
classical MPS format of commercial mathematical programming systems
\cite{MPS} with 12-character fields for all data values.

\textit{\textbf{The pilot problems.}}
These are from a set of economic models developed by Professor
George Dantzig's group in the Systems Optimization Laboratory at
Stanford University during the 1980s.  They are available from Netlib
\cite{lpdata} and have been used in previous computational
studies (e.g., \cite{Koch2003final}).  We use three
examples of increasing size: pilot4, pilot, pilot87.
In Table~\ref{tab:results}, three lines for each problem show
the results of steps D, Q1, Q2 of the DQQ procedure.

\begin{table}[p] 
  \caption{\small Three pilot models from Netlib \cite{lpdata}, and 
    eight \Meszaros{} \emph{problematic} problems \cite{MeszarosLP}.
    Dimensions of $m \times n$ constraint matrices $S$, 
    and size of the largest optimal primal and dual variables
    $v^*$, $y^*$.}
\label{tab:dims}
\centering
\bigskip
\begin{footnotesize}
\begin{tabular}{|@{\,}l@{\ }|@{\,\,}r@{\,\,}r@{\,\,\,\,}r@{\,\,\,\,\,\,}r@{\,\,}|@{\,\,}r@{\,\,\,}r@{\,}|}
\hline
   model    & $m\ \ $ &$n\ \ \ $&\!\!\!nnz($S$) &$\!\!\!\!\max |S_{ij}|$&
                                 $\norm{v^*}_\infty\,\,$&$\norm{y^*}_\infty\,\,$
\\ \hline
   pilot4   &     411 &   1000 &     5145 & 2.8\E+04 & 9.6\E+04 & 2.7\E+02
\\ pilot    &    1442 &   3652 &    43220 & 1.5\E+02 & 4.1\E+03 & 2.0\E+02
\\ pilot87  &    2031 &   4883 &    73804 & 1.0\E+03 & 2.4\E+04 & 1.1\E+01
\\ \hline
   de063155 &     853 &   1488 &     5405 & 8.3\E+11 & 3.1\E+13 & 6.2\E+04
\\ de063157 &     937 &   1488 &     5551 & 2.3\E+18 & 2.3\E+17 & 6.2\E+04
\\ de080285 &     937 &   1488 &     5471 & 9.7\E+02 & 1.1\E+02 & 2.6\E+01
\\ gen1     &     770 &   2560 &    64621 & 1.0\E+00 & 3.0\E+00 & 1.0\E+00
\\ gen2     &    1122 &   3264 &    84095 & 1.0\E+00 & 3.3\E+00 & 1.0\E+00
\\ gen4     &    1538 &   4297 &   110174 & 1.0\E+00 & 3.0\E+00 & 1.0\E+00
\\ l30      &    2702 &  15380 &    64790 & 1.8\E+00 & 1.0\E+09 & 4.2\E+00
\\ iprob    &    3002 &   3001 &    12000 & 9.9\E+03 & 3.1\E+02 & 1.1\E+00
\\ \hline
\end{tabular}
\end{footnotesize}
\end{table}

\begin{table}[p] 
  \caption{\small 
           Iterations and runtimes in seconds for step~D (Double MINOS)
           and steps~\hbox{Q1, Q2} (Quad MINOS)
           on the problems of Table~\ref{tab:dims}.
           Pinf and Dinf = final maximum primal and dual infeasibilities
           ($\log_{10}$ values tabulated, except \Bzz{} means 0).
           Problem iprob is infeasible.
           Bold figures show Pinf and Dinf at the end of step Q2.
           Note that $\text{Pinf}/\norm{v^*}_\infty$ and
           $\text{Dinf}/\norm{y^*}_\infty$ are all $O(10^{-30})$
           or smaller, even though only $O(10^{-15})$ was requested.
           This is an unexpectedly favorable empirical finding.
}
\label{tab:results}
\centering
\bigskip
\begin{footnotesize}
\begin{tabular}{|@{\,}l@{\,}|@{\,}r@{\,}r@{\,\,}|r|@{\;}r@{\ \;\,}r@{\,}|}
\hline      
   model    &  Itns &  Times & Final objective \quad\     &     Pinf &  Dinf
\\ \hline
   pilot4   &  1464 &    0.1 &$-$2.5811392619\E+03        &    $-$05 & $-$12
\\          &     7 &    0.0 &$-$2.5811392589\E+03        &    $-$52 & $-$31 
\\          &     0 &    0.0 &$-$2.5811392589\E+03        &     \Bzz & $-$\B29
\\ pilot    & 16060 &    9.0 &$-$5.5739887685\E+02        &    $-$06 & $-$03 
\\          &    29 &    0.3 &$-$5.5748972928\E+02        &       -- & $-$32 
\\          &     0 &    0.1 &$-$5.5748972928\E+02        &     \Bzz & $-$\B32
\\ pilot87  & 19340 &   22.6 &   3.0171038489\E+02        &    $-$08 & $-$06
\\          &    32 &    0.9 &   3.0171034733\E+02        &       -- & $-$32 
\\          &     0 &    0.6 &   3.0171034733\E+02        &     \Bzz & $-$\B33
\\ \hline
   de063155 &   973 &    0.1 &   1.8968895791\E+10        &    $-$14 & $+$03
\\          &    90 &    0.1 &   9.8830944565\E+09        &       -- & $-$27
\\          &     0 &    0.0 &   9.8830944565\E+09        &     \Bzz & $-$\B24
\\ de063157 &  1473 &    0.1 &   2.6170359397\E+12        &       -- & $+$08
\\          &   286 &    0.2 &   2.1528501109\E+07        &    $-$29 & $-$12 
\\          &     0 &    0.0 &   2.1528501109\E+07        &     \Bzz & $-$\B12
\\ de080285 &   418 &    0.0 &   1.4495817688\E+01        &    $-$09 & $-$02
\\          &   132 &    0.1 &   1.3924732864\E+01        &    $-$35 & $-$32
\\          &     0 &    0.0 &   1.3924732864\E+01        &     \Bzz & $-$\B32
\\ gen1     &303212 &  156.9 &$-$8.1861282705\E-08        &    $-$06 & $-$13
\\          &216746 & 3431.2 &   1.2939275026\E-06        &    $-$12 & $-$31
\\          &  8304 &  112.5 &   1.2953925804\E-06        &  $-$\B46 & $-$\B31
\\ gen2     & 45905 &   60.0 &   3.2927907833\E+00        &    $-$04 & $-$12 
\\          &  2192 &  359.9 &   3.2927907840\E+00        &       -- & $-$29 
\\          &     0 &   10.4 &   3.2927907840\E+00        &     \Bzz & $-$\B32
\\ gen4     & 38111 &  151.3 &$-$1.2724113149\E-07        &    $-$07 & $-$12 
\\          & 58118 & 6420.2 &   2.8932557999\E-06        &    $-$12 & $-$31 
\\          &    50 &    4.3 &   2.8933064888\E-06        &  $-$\B53 & $-$\B30
\\ l30      &1302602&  805.6 &   9.5266141670\E-01        &    $-$08 & $-$09
\\          & 500000& 6168.8 &$-$4.5793509329\E-26        &    $-$25 & $-$00
\\          &  16292&  204.4 &$-$6.6656750251\E-26        &  $-$\B25 & $-$\B31
\\ iprob    &  1087 &    0.2 &   2.6891551285\E+03        &    $+$02 & $-$11
\\          &     0 &    0.0 &   2.6891551285\E+03        &    $+$02 & $-$30
\\          &     0 &    0.0 &   2.6891551285\E+03        &    $+$02 & $-$\B28
\\ \hline
\end{tabular}
\end{footnotesize}
\end{table}

Line 1 for pilot shows that Double MINOS with cold start and scaling
(step D) required 16060 simplex iterations and 9 CPU seconds.
The unscaled primal solution $v$ satisfied the constraints in
\eqref{eq:fba} to within $O(10^{-6})$ and the dual solution $y$
satisfied the optimality conditions to within $O(10^{-3})$.

Line 2 for pilot shows that Quad MINOS starting from that point with
scaling (step Q1) needed only 29 iterations and 0.3 seconds to obtain
a very accurate solution.

Line 3 for pilot shows that in the ``insurance'' step Q2, Quad
MINOS warm-starting again but with no scaling gave an equally good
solution (maximum infeasibilities 0.0 and $O(10^{-32})$).

The final Double and Quad objective values differ
in the 4th significant digit, as suggested by removal of step D's
$O(10^{-3})$ dual infeasibility.

Results for the other pilot problems are analogous.

\textit{\textbf{The \Meszaros{} problematic problems.}}  Our DQQ
procedure was initially developed for this set of difficult LO
problems collected by M\'esz\'aros \cite{MeszarosLP}, who names them
\emph{problematic} and notes that ``\emph{modeling mistakes made these
  problems ``crazy,'' but they are excellent examples to test
  numerical robustness of a solver}.''  They were provided as MPS
files \cite{MultiscaleCollaboration}.
The first two problems
have unusually large entries in the constraint matrix $S$.  The step D
objective value for de063155 has at best 1 digit of precision, and is
quite erroneous for de063157.  Nevertheless, the step Q1 and Q2
solutions are seen to be highly accurate (small Pinf and Dinf values)
when the solution norms are taken into account.

The gen problems come from image reconstruction, with no large
entries in $S$, $v$, $y$ but highly degenerate primal solutions $v$.
(In steps D and Q1 for gen1, 60\% of the iterations made no
improvement to the objective, and the final solution has 30\% of the
basic variables on their lower bound.)  For gen1,
step Q1 gave an almost feasible initial solution
(253 basic variables outside their bounds by more than $10^{-15}$ with
a sum of infeasibilities of only $O(10^{-8})$), yet over 200,000
iterations were needed in step Q1 to reach optimality.  These examples
show that Quad precision does not remove the need for a more rigorous
anti-degeneracy procedure (such as Wolfe's method as advocated by
Fletcher \cite{Fletcher2014}), and/or steepest-edge pricing
\cite{ForrestGoldfarb92}, to reduce significantly the total number of
iterations.  Problems gen1 and gen4 show that step Q2 is sometimes
needed to achieve high accuracy.

Problem l30 behaved similarly (80\% degenerate iterations in steps
D and Q1).  The tiny objective value is essentially zero, so we can't
expect the Q1 and Q2 objectives to agree in their leading digits.
The Q1 iterations were inadvertently limited to 500,000, but
step Q2 did not have much further to go.

Problem iprob is an artificial one that was intended to be feasible
with a very ill-conditioned optimal basis, but the MPS file provided
to us contained low-precision data (many entries like 0.604 or
0.0422).  Our Double and Quad runs agree that the problem is
infeasible.  This is an example of Quad removing some doubt that
would be inevitable with just Double.

Table~\ref{tab:results} shows that Quad MINOS almost invariably
achieves {far more accurate solutions than requested}, in the sense
that the maximum primal and dual infeasibilities are almost always far
smaller than $10^{-15}$.  Thus our procedure for handling the
\emph{problematic} problems is appropriate for the systems biology
M and ME models.  Like the gen problems, the ME models showed
40--60\% degenerate iterations in step~D, but fortunately not so many
total iterations in step Q1 (see Table~\ref{tab:MEtime}).  This is
important for FVA and for ME with nonlinear constraints, where there
are many warm starts.

\textit{\textbf{ME models (FBA with coupling constraints).}}
As coupling constraints are often functions of the organism's growth
rate $\mu$, O'Brien et al.\ \cite{Josh2013} consider growth-rate optimization
nonlinearly with the single $\mu$ as the objective in \eqref{eq:fba}
instead of via a linear biomass objective function. Nonlinear
constraints of the form
\begin{eqnarray}
    {\textstyle v_i \ge \mu \sum_j {v_j} / {k^{\mathrm{eff}}_{i,j}} }
\label{eq:nonlinear2}
\end{eqnarray}
are added to \eqref{eq:fba}, where $v_i,v_j,\mu$ are all variables,
and $k^{\mathrm{eff}}_{i,j}$ is an effective rate constant.
Constraints \eqref{eq:nonlinear2} are linear if $\mu$ is fixed at a
specific value $\mu_k$.  O'Brien et al.\ \cite{Josh2013} employ a
binary search on a discrete set of values within an interval
$[\mu_{\min},\mu_{\max}]$ to find the largest $\mu_k \equiv \mu^*$
that keeps the associated linear problem feasible.  Thus, the
procedure requires reliable solution of a sequence of related LO problems.

\textit{\textbf{Flux Variability Analysis (FVA).}} 
After FBA \eqref{eq:fba} returns an optimal objective value $c\T v^*
\!=\! Z_0$, 
FVA examines how far a particular flux
$v_j$ can vary within the feasible region without changing the optimal
objective significantly (if $\gamma \approx 1$):
\begin{eqnarray} \label{eq:fva}
     \min_v \ \pm v_j
  \text{\ \ s.t.\ \ } Sv = 0, \ \ c\T v \ge \gamma Z_0,
                          \ \ l \le v \le u,
\end{eqnarray}
where $0 < \gamma < 1$. Potentially $2n$ LO problems \eqref{eq:fva} are
solved if all reactions are of interest, with warm starts being used
when $j$ is increased to $j+1$ \cite{Ines2010fva}.

For such a sequence of related problems, warm-starting each problem in
Quad would be simplest (calling a single solver), but warm-starting in
Double and then in Quad could sometimes be more efficient.

\medskip
\noindent
\textbf{Conventional iterative refinement.}
A Double simplex solver would be more reliable with the help of
iterative refinement (Wilkinson \cite{Wilkinson1965}), but we
found this inadequate for the biology models
(see DRR procedure in Supplementary Information).

\medskip
\noindent
\textbf{The zoom strategy.}  A step toward warm-starting interior
methods for optimization was proposed in 
\cite{Zoom} to take advantage of the fact that a low-accuracy solution
$(x_1,y_1)$ for a general problem
\begin{eqnarray} \label{eq:LO}
  \min\ c\T x \text{\ \ s.t.\ \ } Ax=b, \ \ \ell \le x \le u
\end{eqnarray}
can be obtained relatively cheaply when an iterative solver
for linear systems is used to compute each search direction.
(The iterative solver must work harder as the interior method
approaches a solution.)  If $(x_1,y_1)$ has at least some correct
digits, the primal residual $r_1 = b - Ax_1$ will be somewhat small
($\norm{r_1} = O(1/\sigma)$ for some $\sigma \gg 1)$ and the dual
residual $d_1 = c - A\T y_1$ will be comparably small in the elements
associated with the final $B$.  If we define
\begin{eqnarray}
   \begin{array}{ll}
         b_2 = \sigma r_1,           & c_2 = \sigma d_1,
   \\ \ell_2 = \sigma(\ell - x_1),   & u_2 = \sigma(u - x_1),
   \\ x = x_1 + \frac{1}{\sigma}x_2, & y = y_1 + \frac{1}{\sigma}y_2,
   \end{array}
\end{eqnarray}
and note that the problem is equivalent to
\begin{eqnarray}
   \min\ c\T x - y_1^T (Ax - b)
         \text{\ \ s.t.\ \ } Ax=b, \ \ell \le x \le u
\end{eqnarray}
with dual variable $y-y_1$, we see that $x_2$ solves
\begin{eqnarray}
   \min\ c_2\T x_2
         \text{\ \ s.t.\ \ } Ax_2=b_2, \ \ell_2 \le x_2 \le u_2
\end{eqnarray}
with dual variable $y_2$.
Importantly, with $\sigma$ chosen carefully we expect
$(x_2,y_2)$ in this ``\emph{zoomed in}'' problem to be of order 1.
Hence we can solve the problem with the same solver as before
(as solvers use absolute tolerances and assume that $A$ and
the solution are of order 1).
If the computed $(x_2,y_2)$ has at least some digits of accuracy,
the correction $x_1 \leftarrow x_1 + \frac{1}{\sigma}x_2$,
$y_1 \leftarrow y_1 + \frac{1}{\sigma}y_2$ will be more accurate
than before.  The process can be repeated.
%
With repeated zooms (named
\emph{refinement rounds} in
\cite{IterativeRefinement2015,gleixner2015thesis}),
the residuals $(r_1,d_1)$ must be computed with increasingly high precision.
Subject to the expense of using rational arithmetic for this purpose,
\cite{IterativeRefinement2015} gives extensive results for over 1000
challenging problems and shows that exceptional accuracy can be
obtained in reasonable time: only 3 or 4 refinements to achieve
$10^{-50}$ precision, and less than 20 refinements to achieve
$10^{-250}$.  The SoPlex80bit solver \cite{SoplexThesis,SoPlex} is used
for each refinement round with feasibility and optimality
tolerances set to $10^{-9}$.
%
%
In \cite{IterativeRefinement2015} the authors recognize that much 
depends on the robustness of the simplex solver used for the original
problem and each refinement.  The potential difficulties are the same
as in each step of our DRR procedure, 
where Double MINOS is on the brink of failure on the Glc problems
because $B$ is frequently near-singular when it is refactorized
every 100 iterations.
A practical answer for \cite{IterativeRefinement2015} is to use a
more accurate floating-point solver such as Quad MINOS (or Quad
versions of SoPlex or SNOPT \cite{gill2005snopt}) for all 
refinement rounds.

\medskip
\noindent
\textbf{DQQ serves the current purpose.}
In the context of ME models whose non-integer data is accurate to only
4 or 5 digits, we don't need $10^{-50}$ precision.
Tables \ref{tab:MEobj} and \ref{tab:results} show that our DQQ
procedure achieves more accuracy than necessary on all tested examples.
For models where the Double solver is expected to encounter difficulty,
step D can use a reasonable iteration limit.  Step Q1 will perform more
of the total work with greatly improved reliability.  Step Q2 provides
a small but important improvement at negligible cost, ensuring small
residuals for the original (unscaled) problem.

\medskip
\noindent
\textbf{Data and software availability.} 
Double and Quad Fortran 77 implementations of MINOS are included
within the Cobra toolbox \cite{schellenberger2011quantitative}.  MPS
or JSON files for all models discussed are available from
\cite{MultiscaleCollaboration}.
Python code for running Double and Quad MINOS on the BiGG JSON files
is also available from \cite{MultiscaleCollaboration}.

\footnotesize






\subsubsection*{\textsf{\textbf{Acknowledgements}}}
\vspace{-3pt}
We thank Jon Dattorro, Philip Gill, Edward O'Brien, Elizabeth Wong,
and several other colleagues for their valuable help.
Joshua Lerman at UC San Diego provided the model named TMA\_ME here
(originally model\_final\_build\_unscaled.mat) and advised us of the
final objective values obtained by SoPlex and
QSopt\_ex.  
Yuekai Sun at Stanford University created the reformulated version
named GlcAlift here.  Ed Klotz of IBM in Carson City NV provided MPS
files for the \Meszaros{} \emph{problematic} problems and lengthy
discussions of their properties.

  This work was supported by
   the National Institute of General Medical Sciences of
   the National Institutes of Health [award U01GM102098]
  and
   the US Department of Energy [award DE-SC0010429].
  The content is solely the responsibility of the authors and does not
  necessarily represent the official views of the funding agencies.

\subsubsection*{\textsf{\textbf{Author Contributions}}}
\vspace{-3pt}
R.F. and M.S. conceived this study.
D.M. and M.S. developed the DQQ procedure and designed the manuscript.
M.S.   developed the Double and Quad MINOS solvers.
L.Y.   implemented Python interfaces and verified the solvers on
       linear and nonlinear ME models.
R.F. implemented Matlab interfaces within the COBRA toolbox.
I.T.   highlighted the impact of coupling constraints in ME models
       and built the largest example, GlcAerWT.
All authors read and revised the manuscript.

\subsubsection*{\textsf{\textbf{Competing Financial Interests}}}
\vspace{-3pt}
The authors declare no competing financial interests.

\noindent
\hbox to 0.47\textwidth{\hrulefill}

\renewcommand{\refname}{\textsf{\textbf{\large References}}}
\bibliographystyle{unsrt}




\newpage
\normalsize
\begin{center}
\section*{\textsf{\textbf{Supplementary Information}}}
\end{center}

\subsection*{\textsf{\textbf{DQQ procedure}}}
\vspace{-6pt}

\begin{figure}[h!]   
\centering
\includegraphics[width=\columnwidth]{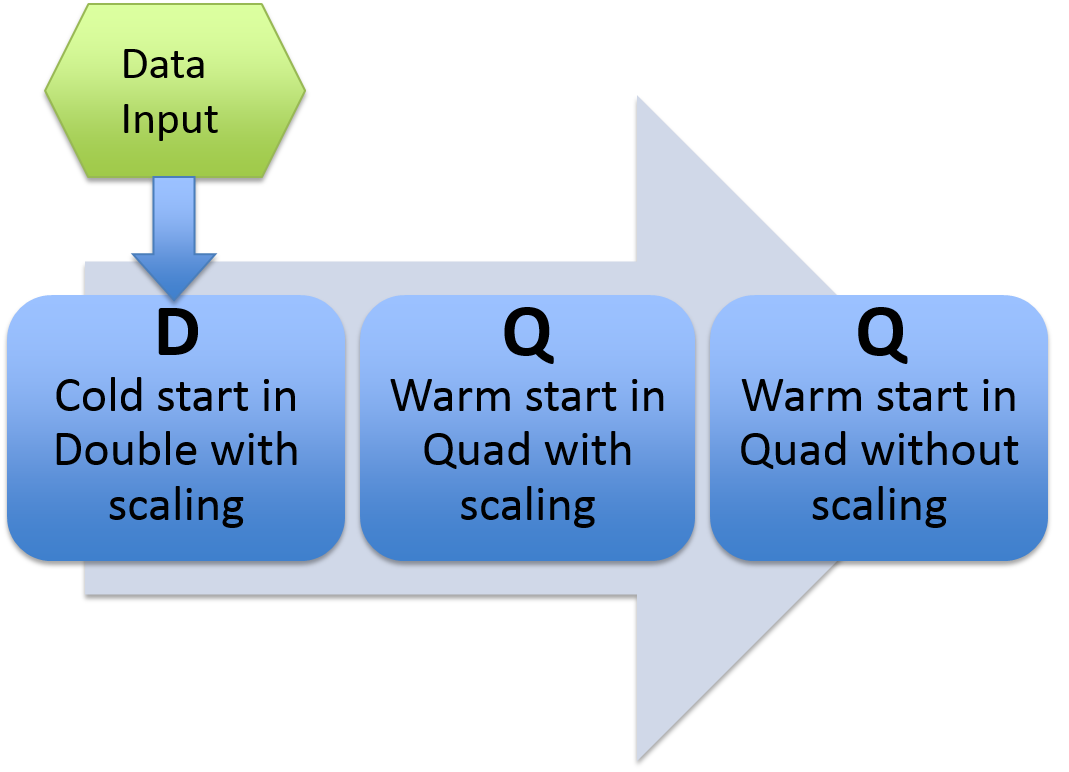}
\vspace*{-15pt}
\caption{\small Flowchart for the 3-step DQQ procedure. \label{fig:3step}}
\end{figure}

Figure \ref{fig:3step} summarizes our approach to achieving
reliability and efficiency for multiscale linear and nonlinear
optimization problems.

The main paper reports application of DQQ to three large ME models
(TMA\_ME, GlcAerWT, GlcAlift)
and to some other challenging linear optimization problems (the pilot
economic models and the \Meszaros{} ``problematic'' set).
Below we provide the following supplementary information:
\begin{itemize}
\item Solution of 78 Metabolic models by Double and Quad MINOS,
      verifying that the Double solver gives reliable results.

\item Solution of two slightly different forms of the TMA\_ME model,
      showing robustness of solution values with respect to $O(10^{-6})$
      relative perturbations of the data.

\item Some details of the Double and Quad MINOS implementations.

\item Experiments with conventional iterative refinement (DRR procedure).

\item Results with Gurobi on the ME models.

\item Results with SoPlex80bit on the ME models.
\end{itemize}

\subsection*{\textsf{\textbf{Metabolic models with Quad solvers admit biomass synthesis}}}
\vspace{-6pt}

COBRA models of metabolic networks assume the existence of at least
one steady-state flux vector that satisfies the imposed constraints
and admits a non-zero optimal objective. Where the objective is to
maximize a biomass synthesis reaction, the corresponding FBA problem
should admit a nonzero biomass synthesis rate. It is established
practice to solve monoscale metabolic FBA problems with Double
solvers, so one may ask: do biomass synthesis predictions from
metabolic models hold when higher precision solvers are applied to the
same FBA problem? We tested 78 M models derived from the BiGG
database \cite{KingLu2015} using Double and Quad MINOS. We
downloaded these models in the JSON format and parsed them using the
JSON reader in cobrapy \cite{cobrapyBMC}. The models were not modified
after loading, so all constraints, bounds, and objective coefficients
were used as in the original files.  All models were feasible using
both Double and Quad, and all but five models had an optimal objective
value greater than zero. Of these five models, four simply had
all-zero objective coefficients, while the remaining (RECON1) model
maximized a single reaction (S6T14g) but its optimal value was zero.
The maximum difference in objective value between Double and Quad was
$2.6 \times 10^{-12}$.
The additional precision provided by Quad MINOS enabled us to conclude
efficiently and effectively that the 78 metabolic models could be
solved reliably using a Double solver. This conclusion is consistent
with previous findings by Ebrahim et al.~\cite{ebrahimStandards2015}.




\subsection*{\textsf{\textbf{Robustness of solution values for TMA\_ME}}}
\vspace{-6pt}

TMA\_ME \cite{Josh2012}
was the first ME model that we used for Quad experiments.  The
data $S$, $c$, $\ell$, $u$  came as a Matlab
structure with $c_j=0$, $\ell_j=0$, $u_j=1000$ for most $j$, except
four variables had smaller upper bounds, the last variable had
moderate positive bounds, and 64 variables were fixed at zero.  The
objective was to maximize flux $v_{17533}$. We output the data to a
plain text file. 
%
%
Most entries of $S$ were integers (represented exactly), but about
5000 $S_{ij}$ values were of the form
   8.037943687315\E-01 or 3.488862338191\E-06
with 13 significant digits.  The text data was read into Double and
Quad versions of a proto\-type Fortran 90 implementation of SQOPT
\cite{gill2005snopt}.

For the present work, we used the same Matlab data to generate an MPS
file for input into MINOS.  Since this is limited to 
6 significant digits, the values in the preceding paragraph were
rounded to 8.03794\E-01 and 3.48886\E-06 and in total about 5000
$S_{ij}$ values had $O(10^{-6})$ relative perturbations of this kind.
This was a fortuitous limitation for the ME models. We have been
concerned that such data perturbations could alter the FBA solution
greatly because the final basis matrices could have condition number
as large as $10^6$ or even $10^{12}$ (as estimated by LUSOL
\cite{LUSOL} each time SQOPT or MINOS factorizes the current basis
$B$).  However, in comparing Quad SQOPT and Quad MINOS with SoPlex
\cite{SoplexThesis,SoPlex} and the exact simplex solver QSopt\_ex
\cite{QSopt_ex}, we observe in Table~\ref{tab:Lerman} that the final
objective values for TMA\_ME in Matlab data reported by QSopt\_ex and
Quad SQOPT match in every digit.  Moreover, the objective value
achieved by Quad MINOS on the perturbed data in MPS format agrees to 5
digits of the results from the exact solver QSopt\_ex on the
``accurate" data. These results show the robustness of the TMA\_ME
model and our 34-digit Quad solvers.

\begin{table}[t] 
  \caption{\small TMA\_ME model.
    Robustness of objective values computed by four high-accuracy solvers
    for two slightly different versions of the problem
    with 13-digit and 6-digit data (from Matlab and MPS data respectively).}
  \label{tab:Lerman}
\vspace*{-5pt}
\begin{center}
\begin{small}
  \begin{tabular}{@{}l@{\,\,}|@{\,\,}c|@{\,\,}l@{}}
     \hline
                  & Optimal objective  &
  \\ \hline
     SoPlex80bit  & 8.703671403\E-07   & Matlab data
  \\ QSopt\_ex    & 8.703646169\E-07   & Matlab data
  \\ Quad SQOPT   & 8.703646169\E-07   & Matlab data 
  \\ Quad MINOS   & 8.703631539\E-07   & MPS data    
  \\ \hline
  \end{tabular}
\end{small}
\end{center}
\vspace*{-14pt}
\end{table}

\begin{table}[t] 
  \caption{\small TMA\_ME model.
    Robustness of small solution values $v_j$ and $w_j$ computed by Quad MINOS
    for two slightly different versions
    (Matlab and MPS data respectively).}
  \label{tab:somevw}
\begin{center}
\begin{small}
  \begin{tabular}{@{}l@{\,\,}|@{\,\,}c@{\,\,}c@{\,\,}c@{}}
     \hline
     \ $j$ &     107       &      201      &     302       
  \\ \hline    
     $v_j$ & 2.336815\E-06 & 8.703646\E-07 & 1.454536\E-11  
  \\           
      $w_j$ & 2.336823\E-06 & 8.703632\E-07 & 1.454540\E-11 
  \\ \hline
  \end{tabular}
\end{small}
\end{center}
\vspace*{-14pt}
\end{table}

\begin{table}[t] 
  \caption{\small TMA\_ME model.
    The values of 9 fluxes $v_j, w_j$ computed by Quad MINOS
    for two slightly different versions of the problem, revealing robustness
    of all 9 solution pairs.  These values have 1 digit of agreement.
    Almost all 17535 pairs of values agree to 5 or more digits.}
  \label{tab:vw}
\begin{center}
\begin{small}
   \begin{tabular}{@{}c@{\,\,}|@{\,\,\,\,}c@{\,\,\,\,}c@{\,\,}c@{}}
   \hline
       $j$  &  $v_j$  & $w_j$  & Relative difference
   \\ \hline
      16383 &  6.07\E-07 &  2.04\E-06 &  0.70
   \\ 16459 &  1.71\E-06 &  2.18\E-06 &  0.22
   \\ 16483 &  2.47\E-06 &  5.99\E-07 &  0.76
   \\ 16730 &  1.44\E-06 &  7.87\E-07 &  0.46
   \\ 17461 &  1.71\E-06 &  2.18\E-06 &  0.22
   \\ 17462 &  2.47\E-06 &  5.99\E-07 &  0.76
   \\ 17478 &  6.07\E-07 &  2.04\E-06 &  0.70
   \\ 17507 &  1.44\E-06 &  7.87\E-07 &  0.46
   \\ 17517 &  8.70\E-07 &  2.97\E-06 &  0.71
   \\ \hline
   \end{tabular}
\end{small}
\end{center}
\vspace*{-14pt}
\end{table}

More importantly, for the most part \emph{even small solution values}
are perturbed in only the 5th or 6th significant digit.  Let $v$ and
$w$ be the solutions obtained on slightly different data.  Some
example values are given in Table~\ref{tab:somevw}.
Among all $j$ for which $\max(v_j,w_j) > \delta_1 = 10^{-15}$ (the
feasibility tolerance), the largest relative difference $|v_j - w_j| /
\max(v_j,w_j)$ was less than $10^{-5}$ for all but 31 variables.  For
22 of these pairs, either $v_j$ or $w_j$ was primal or dual degenerate
(meaning one of them was zero and there are alternative solutions with
the same objective value).  The remaining 9 variables had
\hbox{$v_j$, $w_j$} values shown in Table~\ref{tab:vw}.

We see that the values are small (the same magnitude as the data
perturbation) but for each of the nine pairs there is about 1 digit of
agreement.  We could expect thousands of small solution
pairs to differ more, yet for almost \emph{all} 17535 pairs
at least 5 digits agree.

These observations about two forms of TMA\_ME are welcome
empirical evidence of the robustness of this multiscale model.  Quad
solvers can be applied to evaluate the robustness of future
(increasingly large) models of metabolic networks by enabling similar
comparison of high-accuracy solutions for slightly different problems.

\subsection*{\textsf{\textbf{MINOS implementation}}}
\vspace{-6pt}

MINOS \cite{MurS78, MurS82} is a linear and nonlinear optimization
solver implemented in Fortran 77 to solve problems of the form
\begin{equation}
   \min_v\ c\T v + \varphi(v)
   \text{\ \ s.t.\ \ } \ell \le \pmat{v \\ Sv \\ f(v)} \le u,
   \label{eq:NLP}
\end{equation} 
%
where $\varphi(v)$ is a smooth nonlinear function
and $f(v)$ is a vector of smooth nonlinear functions.
The matrix $S$ and the Jacobian of $f(v)$ are assumed to be
sparse.

Let Single/Double/Quad denote the floating-point formats
defined in the 2008 IEEE 754 standard \cite{IEEE754}
with about 7/16/34 digits of precision, respectively.  Single is
not useful in the present context, and Double may not ensure adequate
accuracy for multi\-scale problems.  This is the reason for our work.
Since release 4.6 of the GCC C and Fortran compilers \cite{GCC},
Quad has been available via
the {\tt long double} and {\tt real(16)} data types.  Thus, we have
made a Quad version of 
Double MINOS using the GNU gfortran compiler (GNU Fortran 5.2.0).

On today's machines, Double is implemented in hardware, while Quad (if
available) is typically implemented in a software library, in this case
GCC libquadmath \cite{GCC-libquadmath}.

For Double MINOS, floating-point variables are declared {\tt real(8)}
($\approx 16$ digits).  For Quad MINOS, they are {\tt real(16)}
($\approx 34$ digits) with the data $S, c, \ell, u$ stored
in Quad even though they are not known to that precision.  This allows
operations such as $Sv$ and $S\T y$ to be carried out directly on the
elements of $S$ and the Quad vectors $v,y$.  If $S$ were stored in
Double, such products would require each entry $S_{ij}$ to be
converted from Double to Quad at runtime (many times).

The primal simplex solver in MINOS includes geometric mean scaling
\cite{Fourer82a}, the EXPAND anti-degeneracy
procedure \cite{GilMSW89c}, and partial pricing (but no steepest-edge
pricing, which would generally reduce total iterations and time).
Basis LU factorizations and updates are handled by LUSOL \cite{LUSOL}.
Cold starts use a Crash procedure to find a triangular initial basis
matrix.  Basis files are used to preserve solutions between runs
and to enable warm starts.

{Scaling} is commonly applied to linear programs to make the scaled
data and solution values closer to 1.  Feasibility and optimality
tolerances can be chosen more easily for the scaled problem, and LU
factors of the basis matrix are more likely to be sparse.  For
geometric mean scaling, several passes are made through the columns
and rows of $S$ to compute a scale factor for each column and
row.  A difficulty is that the scaled problem may solve to within
specified feasibility and optimality tolerances, but when the solution
is unscaled it may lie significantly outside the original (unscaled)
bounds.

{EXPAND} tries to accommodate consecutive ``degenerate'' simplex
iterations that make no improvement to the objective function.  The
problem bounds are effectively expanded a tiny amount each iteration
to permit nonzero improvement.  Convergence is usually achieved but is
not theoretically guaranteed \cite{HallMcKinnon2004}.
Progress sometimes stalls for long sequences of iterations.

{LUSOL} bounds the subdiagonals of $L$ when the current basis matrix
$B$ is factorized as $P_1 B P_2 = LU$ with some permutations $P_1$,
$P_2$.  It also bounds off-diagonal elements of elementary triangular
factors $L_j$ that update $L$ in product form each simplex iteration.
(The diagonals of $L$ and each $L_j$ are implicitly 1.)  Maximum
numerical stability would be achieved by setting the LU Factor and
Update tolerances to be near 1.0, but larger values are typically
chosen to balance stability with sparsity.  For safety, we specify
1.9 in step D of DQQ. 
This value guards against unstable factorization of the deceptive
matrix $\text{diag}(-1\ 2\ 1)$, and improves the reliability of Double
MINOS in the present context.


\subsection*{\textsf{\textbf{Conventional iterative refinement}}}
\vspace{-6pt}

For the biology models, our aim is to satisfy Feasibility and
Optimality tolerances of $10^{-15}$ (close to Double precision).  It
is reasonable to suppose that this could be achieved within a Double
simplex solver by implementing iterative refinement (Wilkinson
\cite{Wilkinson1965}) for every linear system involving the basis
matrix $B$ or $B\T$.  This is a more sparing use of Quad precision
than our DQQ procedure.
For example, each time the current $B$ is factorized directly
(typically a new  sparse LU factorization every 100 iterations), the
constraints $Sv = 0$ can be satisfied more accurately by computing the
primal residual $r = 0 - Sv$ from the current solution $v$, solving $B
\Delta v_B = r$, and updating $v_B \leftarrow v_B + \Delta v_B$.  In general,
the new $v$ will not be significantly more accurate unless $r$ is
computed in Quad.  (If $B$ is nearly singular, more than one
refinement may be needed.)  Similarly for solving $B\T y = c_B$ after
refactorization, and for two systems of the form $Bp = a$ and
$B\T y = c_B$ each iteration of the simplex method.

\begin{table}[t] 
  \caption{\small DRR procedure on three ME models.
           Iterations and runtimes in seconds for step~D (Double MINOS
           with scaling)
           and steps~\hbox{R1, R2} (Double MINOS with iterative
           refinement, with and without scaling).
           Pinf and Dinf = final maximum primal and dual infeasibilities
           ($\log_{10}$ values tabulated).
           Bold figures show Pinf and Dinf at the end of step R2.
           The fourth line for each model shows the correct objective value
           (from step Q2 of DQQ).}
\label{tab:DRR}
\centering
\bigskip
\begin{footnotesize}
\begin{tabular}{|@{\,}l@{\,}|@{\,}r@{\,}r@{\,\,}|r|@{\;}r@{\ \;\,}r@{\,}|}
\hline
   model    &  Itns &  Times & Final objective \quad\     &     Pinf &  Dinf
\\ \hline
   TMA\_ME  & 21026 &  350.9 &   8.3789966820\E-07        &    $-$06 & $-$05
\\          &   422 &   25.4 &   8.6990918717\E-07        &    $-$08 & $-$07 
\\          &    71 &    0.0 &   8.7035701805\E-07        &  $-$\B10 & $-$\B10
\\          &       &        &   8.7036315385\E-07        &          &
\\ GlcAerWT & 47718 &10567.8 &$-$6.7687059922\E+05        &    $-$04 & $+$00
\\          &   907 & 1442.7 &$-$7.0344344753\E+05        &    $-$04 & $-$04
\\          &   157 &  151.2 &$-$7.0344342883\E+05        &  $-$\B10 & $-$\B02
\\          &       &        &$-$7.0382449681\E+05        &          &
\\ GlcAlift & 19340 &15913.7 &$-$5.3319574961\E+05        &    $-$03 & $-$01
\\          &   447 &  198.8 &$-$7.0331052509\E+05        &    $-$03 & $-$03 
\\          &   460 &    0.6 &$-$7.0330602383\E+05        &  $-$\B06 & $-$\B10
\\          &       &        &$-$7.0434008750\E+05        &          &
\\ \hline
\end{tabular}
\end{footnotesize}
\vspace*{-8pt}
\end{table}


By analogy with DQQ, we implemented the following procedure within a
test version of Double MINOS.  Note that ``iterative refinement'' in
steps R1, R2 means a single refinement for each $B$ or $B\T$ system,
with residuals $-Sv$, $a -Bp$, $c_B-B\T y$ computed in Quad as just
described.

\smallskip

\noindent
\textbf{DRR procedure.}  

\textit{Step D} (Cold start with scaling):
  Apply Double MINOS with moderately strict options.
  Save the final basis.
  
\textit{Step R1} (Warm start with refinement and scaling):
  Start Double MINOS from the saved basis with stricter
  tolerances and iterative refinement.  Save the final basis.
  
\textit{Step R2} (Warm start with refinement but no scaling):
  Start Double MINOS from the second saved basis without scaling but
  with stricter LU tolerances and iterative refinement.

\smallskip
Step D is the same as for DQQ (with no refinement).  The
runtime options for each step are the same as for DQQ,
except in steps R1, R2 the tolerances 1\E-15 were relaxed to 1\E-9.

In Table \ref{tab:DRR} we see that this simplified (cheap) form of
iterative refinement is only partially successful, with step R2 achieving
only 4, 3, and 2 correct digits in the final objective.  For GlcAerWT,
steps R1 and R2 encountered frequent near-singularities in the LU
factors of $B$ (requiring excessive refactorizations and alteration of
$B$), and in step R2, the single refinement could not always achieve
full Double precision accuracy for each system.  Additional
refinements would improve the final Pinf and Dinf, but would not
reduce the excessive factorizations.  We conclude that on the bigger
ME problems, a Double solver is on the brink of failure even with the
aid of conventional (Wilkinson-type) iterative refinement of each
system involving $B$ and $B\T$.  We conclude that our DQQ procedure
is a more expensive but vitally more robust approach.


\subsection*{\textsf{\textbf{Results with NEOS/Gurobi}}}
\vspace{-6pt}

For large linear models, commercial solvers have reached a high peak
of efficiency.  It would be ideal to make use of them to the extent
possible.  For example, their Presolve capability allows most of the
optimization to be performed on a greatly reduced form of any typical
model.

\begin{table} 
  \caption{\small Performance of Gurobi with default options
    on three ME models. 
    Note that ``switch to quad'' 
    means switch to 80-bit floating-point (not to IEEE Quad
    precision).  This did not help GLcAerWT.
    For GlcAlift2, the options were NumericFocus 3,
    no Presolve, and no scaling.}
\label{tab:NEOS}
%
\medskip
\begin{footnotesize}
\begin{tabular}{|lll|}
   \hline
   {TMA\_ME}
           &{Presolve}& {$18209 \times 17535 \rightarrow 2386 \times 2925$}
\\ {Optimal}&Iterations& 1703
\\ 0.5 secs& Objective      & {\tt 9.6318438361E-07}
\\         & True obj       & {\tt 8.7036315385E-07}
\\ \hline 
  {GlcAerWT}
           &{Presolve}& {$68299 \times 76664 \rightarrow 18065 \times 26157$}
\\         & {Warning}  & switch to quad (itns $\approx 14000$)
\\ {Numeric error}
           & Iterations     & 593819
\\3715 secs& Objective      &                {\tt 3.2926249E+07}
\\         & True obj       & \hspace*{-7pt} {\tt -7.0382449681E+05}
\\ \hline 
  {GlcAlift}
           &{Presolve}& {$69528 \times 77893 \rightarrow 18063 \times 26155$}
\\         & {Warning}  & switch to quad (itns $\approx 10000$)
\\ {Optimal}&Iterations& 45947
\\ 109 secs& Objective      & \hspace*{-7pt} {\tt -7.043390954E+05}
\\         & True obj       & \hspace*{-7pt} {\tt -7.0434008750E+05}
\\         & {Warning}      & unscaled primal/dual residuals:
\\         &                & \tt 1.07, 1.22E-06
\\ \hline\hline
  {GlcAlift2}
           &                &
\\ {Optimal}&Iterations& 128596
\\ 844 secs& Objective      & \hspace*{-7pt} {\tt {-7.0434}15774E+05}
\\         & True obj       & \hspace*{-7pt} {\tt  -7.0434008750E+05}
\\         & {Warning}      & unscaled primal residual:
\\         &                & \tt 1.05E-05
\\ \hline
\end{tabular}
\end{footnotesize}
\end{table}

Table \ref{tab:NEOS} summarizes the performance of Gurobi
\cite{Gurobi} on three large ME models via the NEOS server
\cite{NEOS}.
The first three results used Gurobi's default runtime options, including
Presolve, Dual simplex, and Scaling (with default FeasibilityTol =
OptimalityTol = $1\E-6$).  TMA\_ME seemed to solve
successfully, but from the Quad MINOS solution we know that Gurobi's
final objective value has no correct digits.  GlcAerWT failed with
``Numeric error'' after many expensive iterations using 80-bit
floating-point.  GlcAlift also switched to 80-bit floating-point.
The scaled problem seemed to solve successfully, but unscaling
damaged the primal residual and this casts significant doubt on the
final solution.  (This is the reason for our research.)

For GlcAlift2 we specified NumericFocus 3 with no Presolve and no
scaling.  These options are appropriate for lifted models
\cite{SunFTS2013}.  Gurobi did not switch to 80-bit arithmetic, yet
achieved 5 correct digits in the objective.  This helps confirm the
value of the lifting strategy of \cite{SunFTS2013}, and would provide
a good starting point for steps Q1 and Q2 of DQQ.
However, DQQ permits us to solve the original model GlcAerWT directly
(without the lifting transformation).

\subsection*{\textsf{\textbf{Results with NEOS/SoPlex80bit}}}
\vspace{-6pt}

Table \ref{tab:NEOS2} summarizes the performance of SoPlex80bit
\cite{SoPlex} on the three ME models via NEOS
with default options.

\begin{table} 
  \caption{\small Performance of SoPlex80bit with default options
           on three ME models.} 
\label{tab:NEOS2}
\medskip
\begin{footnotesize}
\begin{tabular}{|lll|}
   \hline
   {TMA\_ME}
           & Simplifier & {$18209 \times 17535 \rightarrow 10865 \times 11272$}
\\ Optimal & Iterations & 12292
\\ 15 secs & Objective  & {\tt 7.49100071E-07}
\\         & True obj   & {\tt 8.7036315385E-07}
\\ \hline 
  {GlcAerWT} 
           & Simplifier& {$68299 \times 76664 \rightarrow 63897 \times 71008$}
\\ {Optimal}
           & Iterations     & 74526
\\ 765 secs& Objective      & \hspace*{-7pt} {\tt -7.03824497E+05}
\\         & True obj       & \hspace*{-7pt} {\tt -7.0382449681E+05}
\\ \hline 
  {GlcAlift}
           &                & {$69528 \times 77893$}
\\ {Infeasible}
           & Iterations     & 95203
\\1010 secs& Objective      &                {\tt  1.40859732E+11}
\\         & True obj       & \hspace*{-7pt} {\tt -7.0434008750E+05}
\\ \hline
\end{tabular}
\end{footnotesize}
\end{table}

SoPlex80bit was very efficient on all ME models.  For TMA\_ME
the final objective value has the right order of magnitude
but 0 correct digits (1 less than Double MINOS).
For GlcAerWT, the most difficult case,
the final objective is correct in all 9 digits reported.
Note that we could not judge this admirable performance without
our results from DQQ.

The solver's Simplifier reduced the size of the first two models
significantly.  Fortuitously it was not activated on GlcAlift, as if
SoPlex knew that Presolve should not be used with lifted models.
However, the final status and objective value (``problem is solved
[infeasible]'' and 1.4e+11) were incorrect.
This result (like Gurobi on GlcAerWT) emphasizes the importance of
DQQ achieving 20 or more correct digits in all cases.

\subsection*{\textsf{\textbf{Looking ahead}}}
\vspace{-6pt}

The large-scale optimizer SNOPT \cite{gill2005snopt} is maintained as
a Fortran 77 solver \texttt{snopt7} \cite{UCSDsoftware} suitable for
step D of the DQQ procedure.  An accompanying Fortran 2003 version
\texttt{snopt9} has also been developed, for which Double and Quad
libraries can be built with only one line of source code changed.
They are ideal for applying DQQ to future multiscale linear and
nonlinear optimization models in numerous fields.


\small

\renewcommand{\refname}{\textsf{\textbf{\large References}}}
\bibliographystyle{unsrt}

\end{document}